# Electronic bottleneck suppression in next-generation networks with integrated photonic digital-to-analog converters

**Jiawei Meng[1], Mario Miscuglio[1], Jonathan George[1], Aydin Babakhani[2], Volker J. Sorger[1],***

[1]Department of Electrical and Computer Engineering, George Washington University, Washington, DC 20052, USA
[2]Department of Electrical and Computer Engineering, University of California Los Angeles, Los Angeles, CA, 90095, USA

*Corresponding Author: *sorger@gwu.edu*

**Digital-to-analog converters (DAC) are indispensable functional units in signal processing instrumentation and wide-band telecommunication links for both civil and military applications. Since photonic systems are capable of high data throughput and low latency, an increasingly found system limitation stems from the required domain-crossing such as digital-to-analog, and electronic-to-optical. A photonic DAC implementation, in contrast, enables a seamless signal conversion with respect to both energy efficiency and short signal delay, often require bulky discrete optical components and electric-optic transformation hence introducing inefficiencies. Here, we introduce a novel coherent parallel photonic DAC concept along with an experimental demonstration capable of performing this digital-to-analog conversion without optic-electric-optic domain crossing. This design hence guarantees a linear intensity weighting among bits operating at high sampling rates, yet at a reduced footprint and power consumption compared to other photonic alternatives. Importantly, this photonic DAC could create seamless interfaces of next-generation data processing hardware for data-centers, task-specific compute accelerators such as neuromorphic engines, and network edge processing applications.**

**Introduction**
The total annual global IP traffic is estimated to reach 4.8 ZB per year by 2022[1]. The continuous increase in demand for both low-latency access and efficient processing of data demands innovative platforms to perform computational tasks closer to the edge of the network. If successful, this will enable next-generation networks to analyze important data efficiently and in near real-time. In this context, most of the world's data already propagates in optical fibers, which support both high channel rates and throughput. Yet, the network's bottlenecks in terms of power consumption and throughput are found in limitations arising from connections and interfaces at its edge; that is, peripheral input/output (I/O) devices such as digital systems or sensors require a digital-to-analog conversion (DAC), and vice versa (ADC). Therefore, it becomes a pressing challenge, especially for large-scale data centers, to optimize or even re-consider network designs to meet the future needs associated with requirements for large dataset-processing and low latency data-access, without sacrificing performance.

Photonic integrated circuits (PIC) have shown the potential to satisfy the demand for high data-processing capabilities while acting on optical data, interfacing to digital systems, and performing both in a compact footprint, with both low latency and power consumption[2]. However, the performance gains of photonic platforms, when interfacing with electrical digital architectures, are limited by the interfaces to/from electronics. Indeed, these interfaces are often limited by the achievable bandwidth as well as the resolution of the DACs (or ADCs) in addition to cumbersome domain-crossings between electronics (E) and optics (O).

The ultimate delay- and throughput performance and power consumption of both digital (and analog-digital hybrid) signal processing systems is constrained by CMOS technology, which is fast approaching its fundamental physical limit.[3] For example, a DAC for software-defined transmitters or as an interface to (emerging) computing systems, should be able to process tens of billions of DAC-operations per second with high linearity and high-timing stability.[5] While integrated electronics is capable of providing high accuracy, linear conversion and stability, it nevertheless is intrinsically limited by bandwidth and high timing jitter, since these implementations are based on multi-ladder voltage/current weighting circuits switch-and-latch architectures, or a combination of segmented architectures in addition of binary weighting[4]. However, fabrication-induced delays in sub-micron CMOS architectures, due to fluctuations in the high resistance of the interconnection wires and/or parasitic capacitance, can seriously compromise performance at the high frequencies. Very high sampling rate converters are comprised of multiple data converters with analog circuitry, such as time or analog bandwidth interleaving and multiplexed DACs.[5]

Here we argue for photonic-based DACs to augment optical and electro-optic hybrid networks and information processing system due to their seamless integration and high performance as discussed below. However, since the bandwidth of state-of-the-art high-ENOB (effective number of bits) electronic DACs is lower than that of typical electro-optic modulators, which are inherently immune to electromagnetic interference, photonic DAC solutions should simultaneously enable high sampling rates and conversion efficiency, while being less affected by jitter nor electromagnetic noise. Moreover, these photonic DACs (and ADCs) then allow bypassing the parasitic opto-electric-opto O-E-O conversions, thus facilitating network simplicity and possibly cascadability to other photonic systems such as networks, processors, etc. (**Fig. 1**). Since photonic DACs are intrinsically compatible with optical fiber communication systems (via grating-to-chip alignment processes), they could also be used for low-latency label data routing in miniaturized switching networks or provide interfaces to information preprocessors and classifiers at the edge of the network. Several optics-based DACs have been explored, however mostly using individually-packaged fiber-optic components, which limits their utilization (footprint, power, robustness, ultimate cost); these have generally been based on a variety of different schemes including optical intensity weighting of multiwavelength signal modulated using micro-ring resonators (PIC-based)[6], nonlinear optical loop mirrors[7], or interferometry and polarization multiplexing[8], or rely on phase shifters[9]. In this work we highlight the need to overcome the bandwidth limitations and timing jitter of current electronic-only DACs while also seeking to overcome cumbersome O-E-O conversions[10]. Indeed, the ability to improve energy consumption and reduce latency is of considerable interest, while not sacrificing chip die real-estate, together making a case for a coherent parallel photonic binary-weighted (PBW) DAC as introduced here. PBW-DACs could be essential devices for the realization of the next-generation networks and used at the interfaces of network-edge photonic dedicated devices, in the information 'fog', such as digitally controlled transmitters and receivers or photonic computing architectures, which can significantly lower the cost of running a network by providing edge-cloud capabilities (**Fig.1**). At higher levels in the network, the DAC can be used as digitally controlled photonic micro datacenters and routers for intelligent re-direction of the traffic and label processing.

More recently, a number of different photonic DAC implementations have been proposed, which can be categorized as serial or parallel according to their operating scheme (**Fig. 2**); serial optical DACs (**Fig. 2a**) are usually based on the summation of weighted multiwavelength pulses opportunely spaced in time by properly setting the wavelength spacing and the length of a dispersive medium, which recreates an analog waveform after being detected by a photoreceiver, enabling fast digital to analog conversion. This type can be straightforwardly cascaded for long-distance optical communication systems, which primarily operate in a serial mode. However, their sampling rates are hindered by multiple factors, such as pulse source and stability, dispersion control, the optical modulator, and any dispersion compensator. These serial optical DACs typically trade off bit resolution for sampling rate. Experimental studies show 4-bit serial prototypes with operating speed of 12.5Gs/s[11].

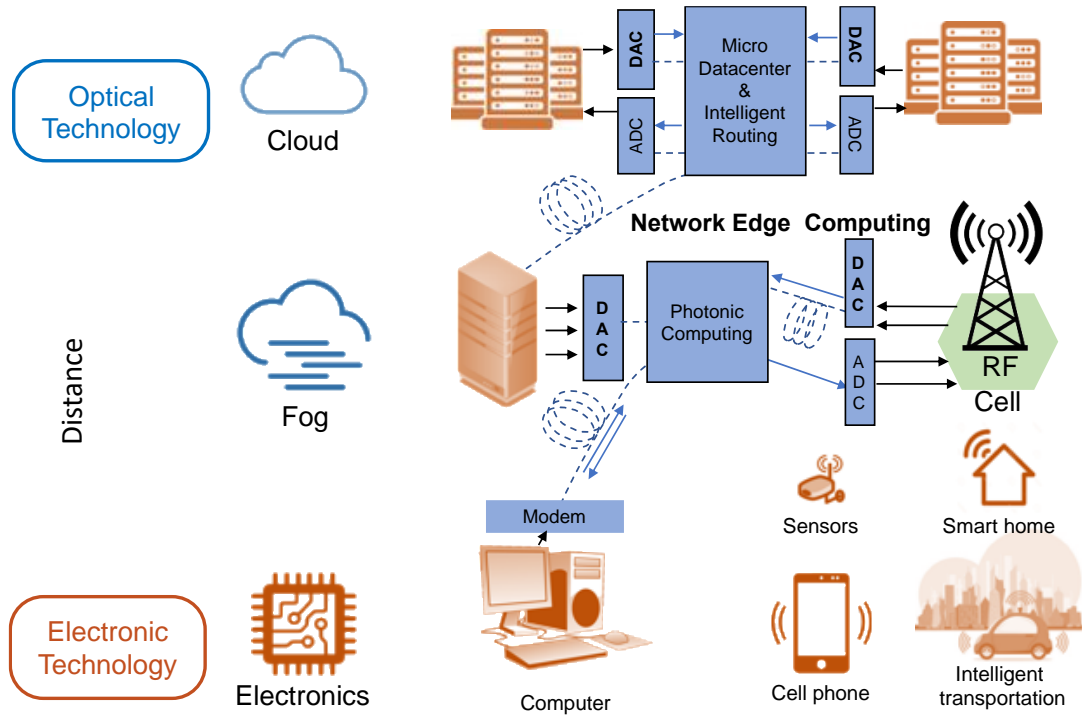

*Figure 1. Schematic representation of the impact and potential uses of a photonic digital-to-analog converter (DAC) in a 5G network. The photonic DAC would be used at the interface between the electronic- and photonic platforms both in the information 'fog' and 'cloud' layers, such as in optical information processing, intelligent routing, and label-date processing or sensor preprocessing at the edge of the network and within servers in the cloud.*

On the other hand, optical DAC implementations are generically characterized by a simpler architecture and usually employ electro-optic modulators to weight the intensities of multiple optical carriers according to an electrical digital signal input where the conversion to an analog signal is performed by a subsequent summation at the end of optical link via a photodetector (**Fig. 2b**).[12,13] However, parallel schemes bear a significant potential towards realizing reduced power consumption, while simultaneously taking full advantage of the combined fast sampling rate and high-bit resolution provided by the multiple parallel channels, as further detailed below (**Fig. 2c**). One limitation, however, associated with this approach, comes from the summation of the modulated optical carriers arising from the requirement that the optical signals must be added coherently to optimize the dynamic range and linear operation of these optical parallel DACs. The issue of incoherent summation is primarily addressed by the use of photodetectors, which integrate the optical power and additional electronics, however limits the operating bandwidth and also adds conversion latency. In general, the conversion to the electrical domain by means of a photodetector is not necessarily desirable, especially for those applications which would still benefit from keeping the analog signal in the optical domain, such as optical machine learning[14-16] or optical telecommunication[17,18], which contributed to the motivation of the coherent parallel PBW DAC introduced here (**Fig.2c**).

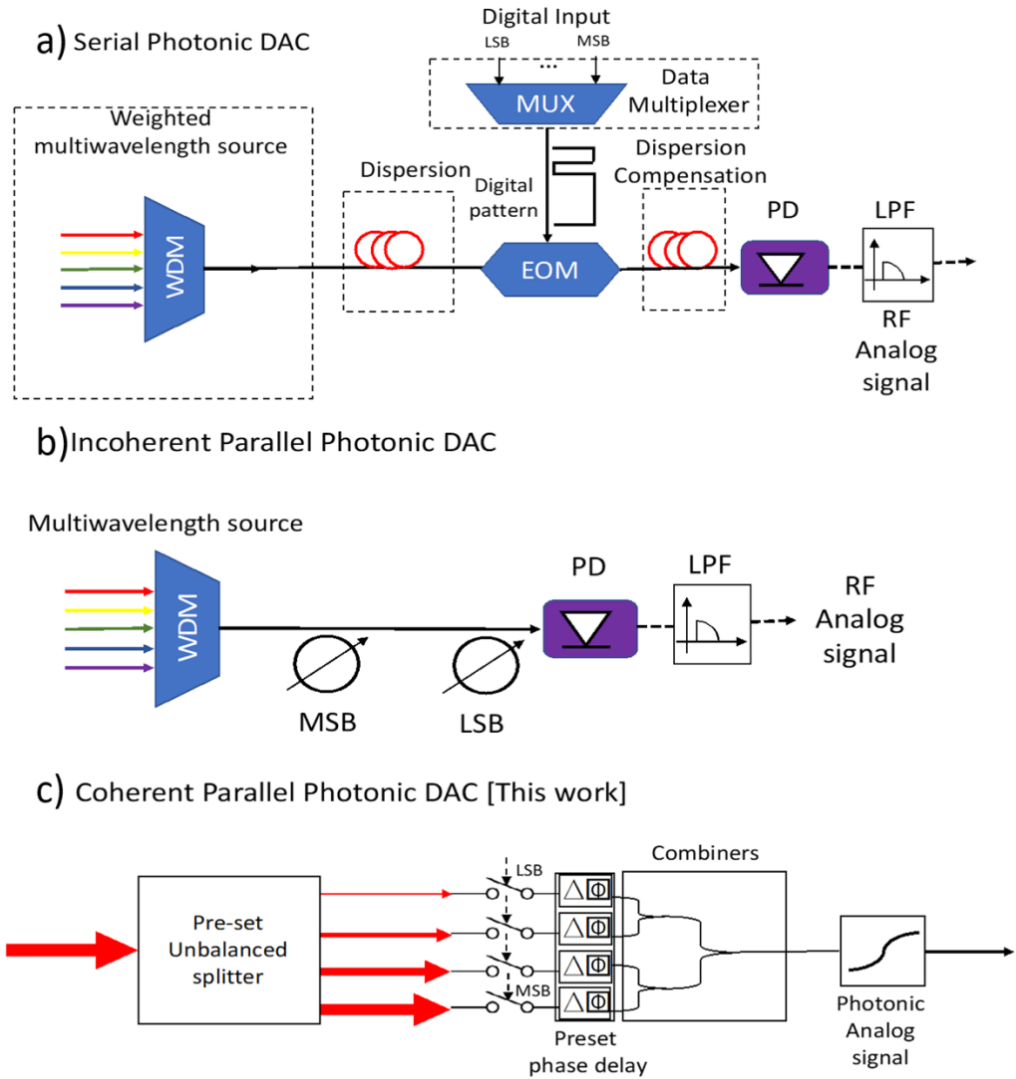

*Figure 2.* Schematic representation of three different implementations of optical DACs, namely serial, incoherent and coherent parallel. *a)* The serial scheme is based on summation of weighted multiwavelength pulses opportunely spaced in time by properly setting the wavelength spacing and the length of dispersive medium. *b)* The parallel implementation is based on weighted integration of multiple wavelengths which encode a bit sequence. *c)* Coherent parallel optical DAC, this work, uses pre-set unbalanced directional couplers that fans-out the optical carrier unevenly into different channels, which are then individually modulated at high-speed using electro-optic modulators (or alternatively 2x2 switches). The pre-determined phase shift (in case of the represented binary '0' or '1') is actively compensated with phase shifters towards a coherent summation using passive waveguide combiners. This enables keeping the signal in the optical domain (a O-E-O conversion is not required) for synergistic use in optical machine learning[14-16] or optical telecommunication[1] systems.

In this work, we propose and demonstrate a PIC-based coherent parallel photonic binary-weighted (PBW) DAC paradigm, which exploits a combination of unbalanced couplers[19] and electro-absorption modulators (**Fig. 2c & 3**). In this configuration, a series of unbalanced couplers divides the optical power into multiple branches in an exponential manner; electro-absorption modulators are employed to absorb the optical power in each branch according to an applied digital electric signal, thus encoding the intensity of the optical signal travelling in each branch only if triggered by a digital input '0'; this would lead also to an alteration of the optical path length in a systematic, hence controllable manner. Nonetheless, systematic phase variations (supplementary online material), if incurred, are compensated by PIC-integrated phase shifters added at each branch, enabling coherent signal summation at the optical output (Y-combiners). As such, this novel approach avoids additional electronics as well as photodetectors. We demonstrate a passive implementation of a 4-bit parallel PBW-DAC and show the potential of conversion speeds in excess of 50 GS/s$_{20}$ along with energy consumptions as low as few pJ/S (S = number of DAC-samples). Further, we analyze the performance degradation of PBW-DAC such as arising from limited extinction ratios of the deployed modulators and discuss the integral and differential nonlinearities of this paradigm. Despite these limitations, we show that we are able to obtain highly linear D/A conversion on an optical signal. Key features of our approach are; a) an analog electrical signal, if desired, can then be obtained via a photodetector; b) this simple and relatively compact paradigm can be employed in numerous ways, for instance in size-weight-area-power (SWAP)-sensitive network-edge data processing enabling; c) low-latency computing or high-speed routing such as for miniaturized data centers, intelligent sensors, and pre-filtering multi-modal data convergence systems.

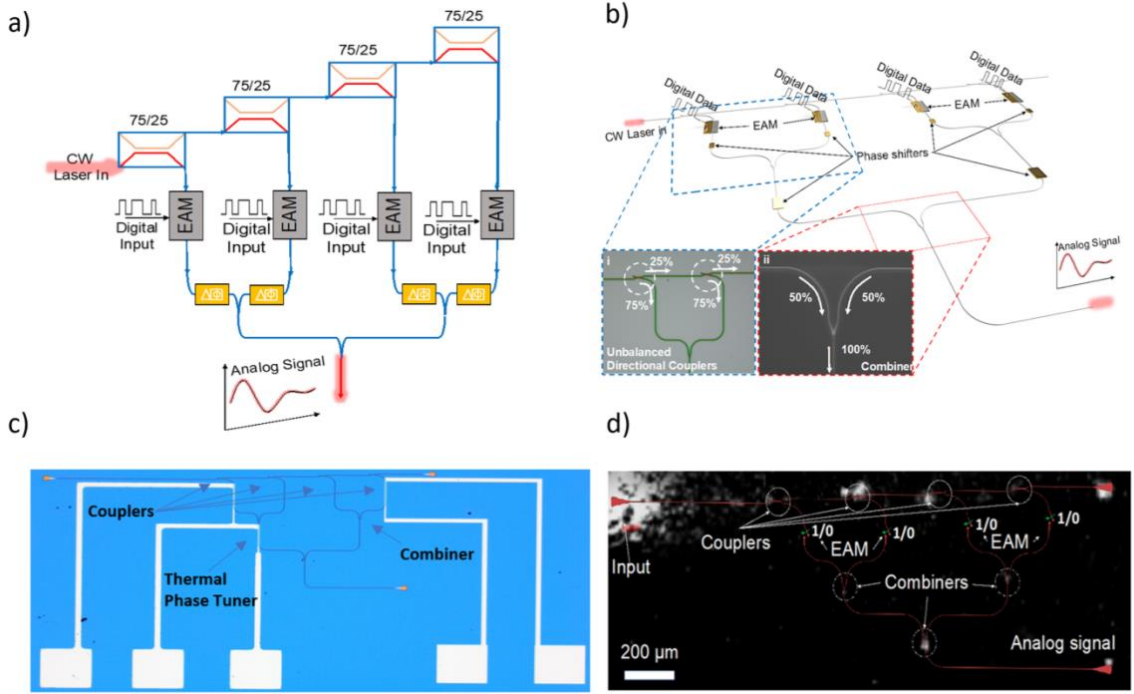

*Figure 3. Schematic diagram of a 4-bit photonic binary-weighted (PBW) DAC based on unbalanced directional couplers and modulators. a) Schematic of the working principle. b) Sketch of a photonic binary-weighted DAC in parallel configuration. A carrier (cw laser) is faned-out into multiple branches via an unbalanced directional coupler (i. SEM image of the directional couplers and its operation) according to the formula: $(1-r)^N$ where r is the splitting ratio (r=0.75) and N is the number of bits. The intensity of the signal is modulated in each branch by an electro-absorption modulator (extinction ratio=4.6dB$_{20}$) according to a digital input electrical signal. In the presence of a binary '0', a systematic phase shift is added to the branch compensating for the optical pathlength variation after modulation. This then allows for the optical signals to be (passively) coherently summed by means of a combiner (ii. SEM image of the combiner and device principle). c) Microscope image of the fabricated passive 4-bit PBW-DAC circuit with the integration of thermal phase shifters for the digital input '1'1'1'1. d) Optical measurements of the passive 4-bit PBW-DAC. The output for each digital input configuration is collected using an IR camera, ultimate with high-speed integrated detectors. The layout of the PIC (red) is superimposed to the IR image*

*for clarity. Shown the optical power for the digital input '1'1'1'1' state.*

Next, we discuss essentials of describing the generic operation of a *N*-bit electro-optical D/A converter utilizing asymmetrical directional couplers for binary-weighting (**Fig. 3b**). The PBW-DAC converts the parallel digital signals comprising of *N*-bits into an analog output signal in the optical domain utilizing a PIC. In this topology a single continuous wave laser is coupled into the PIC where it passes through a sequence of asymmetrical directional couplers with a splitting ratio of 3:1[19,21], which are chosen to optimize the linearity of the output optical power relative to the digital bit inputs (supplementary online material). By design, each consecutive channel $i$, receives a fraction of the optical power ($r = 0.75$) from the previous unbalanced coupling stage. The analog signal power can therefore be written as:

$$E_i = r(1-r)^{N-i} E_{input} \qquad (1)$$

where *E* is the electric field strength. In this way, we obtain *N* separated continuous and weighted waves travelling in *N* channels (i.e. here *N*-waveguide branches, **Fig. 3**), which implements the intensity weighting factors corresponding to each bit of the digital input signals. Thus, the resolution of the PBW-DAC is limited by the number of waveguides *N*. Thanks to the pre-determined and successive splitting obtained by the series of unbalanced couplers and the systematic correction of phase alteration, the signal, modulated according to the *N*-bit sequence, is combined (in-phase) using a sequence of Y-junctions. The final analog optical power output with the combination of each channel is given by:

$$P_{Analog} = \tfrac{1}{2}\sum_{i=2}^{N} \left[ r(1-r)^{N-i} E_{input} \exp\left\{-\tfrac{2\pi}{\lambda}\kappa_{i,eff}^{(0,1)} L\right\} + r(1-r)^{N-i+1} E_{input} \exp\left\{-\tfrac{2\pi}{\lambda}\kappa_{i-1,eff}^{(0,1)} L\right\} \right]^2 \qquad (2)$$

with *i* being even numbers between 2 and *N*, e.g. {2, 4, ..., N} (supplementary online material). The final analog optical signal (Eq. 2) can now be used in the optical domain or be converted, by a photodetector, to obtaining the converted analog signal.

To assess the operation conditions of the PBW-DAC paradigm and gain insights into its performance vectors, the 16 states of the 4-bit BW-DAC prototype demonstration are experimentally validated as individual circuits, where each circuit represents one-bit combination (supplementary online material). For example, when the $i_{th}$ bit is set to a '1', then the signal inside the $i_{th}$ bit-waveguide is kept ON (as supposed to turned OFF) realized by the light 'ON' state of the modulator (or alternatively via a 'BAR-state' of a 2x2 switch) to preserve the optical power passing through the combiners. Contrarily, to emulate the $i_{th}$ bit being equal to a '0', the $i_{th}$ bit-waveguide is disconnected from the output port and thus contributing 'zero' optical power to the output. Notice, that in this way there is a vanishing amount of optical leakage power contributing to the optical analog output signal, thus overestimating the performance. In our experimental demonstration of a 4-bit PBW-DAC, a cw laser source is coupled to the circuit by means of grating couplers and successively split to each arm with sequential weights (**Fig. 3**). To demonstrate the correct functioning for the 4-bit prototype we fabricate and test all 16-states in Si PICs with systematic active tuning of the phase; an exemplary chip and output signal for the '1111' case is depicted in **Figure 3c,d**.

To verify the design functionality and gain further operational insights into the experimental results, we perform numerical simulations (Methods section) including an analysis of i) delay and throughput, ii) frequency and phase stability, and iii) DAC benchmarks such as differential nonlinearity (**Fig. 4**). With PIC yield, repeatability, and timely time-to-market in mind, we use photonic foundry models[19-22]; i.e. modulators have an extinction ratio of 4.6 dB[20] and are driven by non-return-to-zero (NRZ) pulse generator to inject the binary digital bit sequence at 50 Gbit/s (supplementary online material).

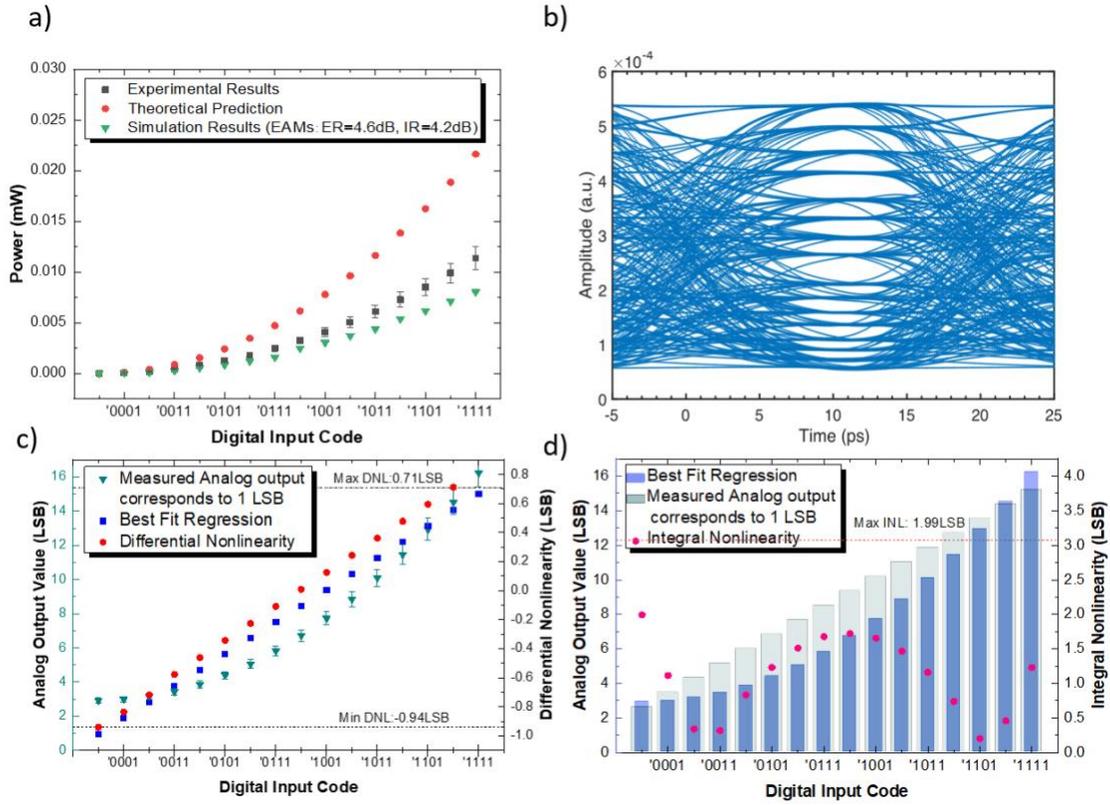

*Figure 4. Performance results of the coherent parallel photonic binary-weighted DAC 4-bit prototype. (a) Analog optical signal output power. Comparison between the experimental PIC results obtained for all $2_N=16$ DAC output states (blue square), the photonic circuit simulated version (green triangle) and the theoretical prediction according to the formula (red circle). The quadratic function is expect since the optical intensity (I) scales with the square of the electric field (E); i.e. $I\sim|E_2|$. (b) Eye diagram of a 4bit DAC assuming thermal noise of the photodetector, static (1ps) and random jitter (1ps) of the pseudo-random bit sequence used as digital input to the modulators. (c) DAC quality performance: Integral nonlinearity (INL) and (d) Differential nonlinearity (DNL) for a given electrical digital signal inputs of the measured analog outputs corresponding to an increment of one (1) least-significant-bit (LSB) with respect to the best fit regression. The DNL is between -0.94 and 0.71 LSB while the INL is smaller than 1.99LSB.*

The measured and simulated output optical powers of the PBW-DAC are in good agreement with each other as highlighted by the (expected) quadric trend, which reflects the relationship between the linear superposition of the electric field waves at the Y-junctions and the resulting intensity (**Eq.2**, **Fig. 4a**). The discrepancies between the model and experiment are associated with reflections at the waveguide representing the different bit-combinations and scattering at the Y-junctions. At the low-power output limit, the PIC-case representing the digital input combination '0000' produces (ideally) zero optical output power. However, the limited dynamic range (i.e. extinction ratio) of the modulators leads to power-leakage in each arm (26% of incident power for each arm considering the EAM with 4.6 dB extinction ratio), thus reducing the dynamic range of the detected DAC output signal when captured by a photodetector. A reduced optical power dynamic range (i.e. signal amplitude change) of the system also impacts the maximum bit-resolution of this DAC paradigm. Furthermore, optoelectronic components bear additional signal noise factors; for instance, the electrical signal driving the modulators is affected by jitter and electrical gaussian noise (Methods). To characterize the output signal quality, we use the signal-to-noise and distortion ratio (SINAD) and obtain 65 and 10 dB, respectively, which is just 2 dB and 1 dB, respectively, lower when compared to the ideal case of zero-power leakage from the modulators, indicating a high accuracy of PBW-DAC (supplementary online material).

Finally, we compare the performance of this PIC-based PBW-DAC to other electronic and optical DAC designs and selecting a medium-high 8-bit resolution (**Table 1**). We use experimental-numerical cross-validation approach from the 4-bit prototype to emulate the 8-bit performance. Focusing on the core-functionality of DACs, we find that sample efficiency of PBW-DAC (>100GS/s) is about 50% higher compared to an integrated electronic DAC at the same speed and about 3x when compared to electronic DACs at twice the speed. The main reason for this is the binary-weighted tree-like architecture in combination with the high efficiency (~10's fJ/bit) and speed (50 Gbps) of electro-optic components such as modulators[22,31], but also synergistic component driver arrays[32]. Note, this estimations also include the required optical laser power to compensated the losses in the PIC, it's wall-plug efficiency (10%) and an 10x SNR safety margin at a (possible) back-end photodetector ($P_{detector-min}$ > 100nW, SNR > 1). The required footprint for PBW-DAC (~1.2mm$^2$) is comparable to integrated electronic solutions. This footprint could be improved further with improved waveguide routing design.

|  | Speed (GS/s) | Power (W) | Sampling Efficiency (GS/J) | Footprint (mm$^2$) | Resolution (#bit) |
|---|---|---|---|---|---|
| Electronic on-chip[27] | 55-65 | 0.75 | 80 | 2.0 | 8 |
| Electronic off-chip[28] | 0.1 | 0.5 | 0.2 | 20.2 | 8 |
| Electronic on-chip[23] | 100 | 2.5 | 40 | 1.6 | 8 |
| **PBW-DAC** [This work] | **50** | **0.45*** | **116** | **~1.2** | **8** |

*Table 1. Speed-power-footprint DAC performance comparison of commercially available electronic DACs with the PBW-DAC introduced here. Where this work's performance is calculated based on the speed of SiGe EAM provided by IMEC, power consumption of 8 EAMs and 6 high speed electro-optic phase shifters for 8-bit PBW-DAC. S = number of digital-to-analog conversions (samples) performed. *Sum of the power consumption of i) the EAMs = 8 times 24mW = 0.19W[22], ii) phase-shifters = 6 times 29mW[31] = 0.17W, iii) two driver arrays for both modulator types consuming 30mW each[32], iv) optical laser power of the photonic DAC circuit to ensure photodetector SNR > 1 ($P_{Detector-min}$>100nW) including 26dB circuit loss using foundry specifications (supplementary online material), 10% laser wall-lug efficiency and 10x SNR safety margin.*

The operating speed and power consumption of the PBW-DAC paradigm is primarily limited by the performance of the component responsible for encoding the digital bit onto the optical domain, namely the modulators. Indeed, the sample speed linearly increases with modulator speed, and future improvements such as recent demonstrations in thin-film LiNbO$_3$ modulators point towards 100 GHz[29] and even 500 GHz[30] performance, which is in stark contrast to electronic circuits (especially digital ones), which are dominated by interconnect delay which does not improve with transistor performance improvements related to scaling.[28] Regarding the maximum resolution, PBW-DAC is limited by the signal discrimination (i.e. extinction ratio) of the modulators; here the optical power of the least-significant-bit (LSB) needs to exceed the leakage power of the most-significant-bit (MSB) for two neighboring binary states (i.e. most sensitive to when the input digital signal for the MSB is '0'). Numerical results predict a possible resolution up to 14-bit when using an absorption modulator with a 4.6-dB extinction ratio (supplementary online information).[26]

Comparing this DAC with other photonic approaches, we find that many optical DAC implementations to-date use non-integrated (off-chip) components and also omit details on power consumption or footprint. Hence, we compare other DAC performance specifications relating to the conversion quality (**Table 2**). We find that the higher DNL and INL values of PBW-DAC are mainly caused by the nonlinear power summation of the serial Y-combiner, which is caused by the optical output intensity being proportional to the square of the electric-field (i.e. Eq.2. supplementary online material). This systematic error can, in principle, be compensated for such as backend using electronic circuitry attached to a photodetector (if converted). However, this will introduced sampling delays, thus rendering the performance gains from the opto-electronic components (e.g. modulators) obsolete. Hence it may be advantages to accept a reduced deviation from the linear regression in lieu of gains in sampling rate.

|  | Speed (GS/s) | DNL (unitless) | INL (unitless) | ENOB (unitless) | Resolution (#bit) |
|---|---|---|---|---|---|
| Parallel Photonic DAC$_{25}$ | 2.5 | 0.5 | 0.5 | 4.1 | 4 |
| Serial Photonic PDAC$_{11}$ | 12.5 | 0.1 | 0.5 | 3.0 | 4 |
| **PBW-DAC** [This work] | **50** | **0.9** | **2.0** | **10.4** | **8** |

*Table 2. Linearity quality and sensitivity comparison of the PBW-DAC with state-of-the-art optical (off-chip) parallel and serial photonic DAC implementations. Differential nonlinearity (DNL), integral nonlinearity (INL) and effective number of bits (ENOB). The reduced linearity of PBW-DAC originates from the coherent summation at the y-combiners where the optical intensity (I) scales with the square of the electric field (E), i.e. $I \sim |E|_2$. See also supplementary online information.*

In conclusion, we introduced, designed, built and tested a coherent parallel binary-weighted photonic integrated circuit-based digital-to-analog-converter (DAC). While just being twice the size compared to electronic chip-based counterparts, this design paradigm synergistically utilizes optical spatial binary-bit-wise parallelism and with non-circuit limiting high sample-rates of opto-electronic components enabling linear sample-rate scaling proportional to the speed of the binary-bit position encoding electro-optic modulator. In this novel photonic DAC the intensity of the optical carriers is split by unbalanced directional couplers, weighted according to multiple input digital signals which drive foundry-ready electro-absorption modulators, and ultimately summed using combiners at the end of the photonic link. The design guarantees a linear intensity weighting consuming as little as 3 pJ/S for a sample-rate of 50 GS/s. We experimentally demonstrate a 4-bit passive prototype of this DAC paradigm and cross-validate experimental results with both full-wave and integrated circuits simulations. In contrast to other parallel photonic DAC implementations, this coherent photonic DAC does not require the signal to be converted in the electrical domain and therefore could support data input/output interfaces of high-throughput applications such as in domain specific compute accelerations, emerging photonic integrated neuromorphic signal processing engines, and network-edge processing and data routing platforms.

## Methods
### Fabrication detail of BW-DAC circuit over SOI platform
Silicon-on-insulator (SOI) based PICs and metal waveguide heaters are fabricated at nanofabrication and imaging center (NIC) at the George Washington University. SOI wafer: Si-epi layer = 220nm, BOX = 2μm. Substrates are diced into 15 mm squares from 150 mm wafers. After Acetone and IPA solvent rinse and hot plate dehydration, the negative photoresist hydrogen silsesquioxane resist (6% HSQ, Dow-Corning XP-1541-006) is spin coated at 4000 rpm and then hotplate baked at 80 C degree for 4 minutes with the coating thickness of 140 ± 10 nm. Electron beam lithography is performed using Raith Voyager system operated at 30 kV energy, 1.38 nA current with the exposure field size of 500 nm. An exposure base dose of 1800 μC/cm$_2$ for the bulk silicon structure and base does of 2700 μC/cm$_2$ for fixed-beam-moving-stage (FBMS) waveguide are used. The resist is developed by dipping the samples into 25% tetramethylammonium hydroxide (TMAH) solvent for 4 minutes, followed by rinsing the chips for 60s in flowing DI water, an IPA rinse for 10 s, and then blown dry with nitrogen. The unexposed silicon area and exposed silicon oxide protection layer were etched by using Apex SLR inductively coupled plasma (ICP) system with 5 sccm gas flow of $SF_6$ and 5 sccm gas flow of $C_4F_8$, pressure of 15 mT, ICP power of 300 W, bias power of 20 W with a platen temperature of 20 C degree. A thick (2000 nm) silicon oxide layer is deposited on the top of chip to increase the refractive index contrast between the silicon layer and cladding towards improving the grating coupler performance using Versaline plasma enhanced chemical vapor deposition (PECVD) system.

**Measurement of the PBW-DAC for different bit pattern**
A 1550 nm laser light, with maximum optical power of 39 mW, is coupled into the PIC through grating couplers and the output signal is captured by a 16-bit resolution IR camera measuring the output optical power density. An electro-optic fiber-coupler based probe station was used to test the various DAC implementations of the 4-bit PBW-DAC prototype circuits. Electrical DC probes were used to control systematically the phase.

**Future improvements**
We note that the overall performance is especially sensitive to the absorption modulators and phase-shifters deployed. While we here considered foundry-ready options, emerging device solutions could be used as well such as those utilizing foundry-near materials such as ITO [33, 34] especially when combined with slow light effects [35] demonstrating very efficient and compact phase-shifters $V\pi L$ = 0.06 V-mm [36, 37] and electro-absorption modulators [38]. Alternatively, two dimensional materials, such as graphene could be deployed for high-speed while preserving high modulation voltage efficiency showing <fJ/bit modulation efficiency [39] and the potential for 100 GHz fast modulation [40]. Scaling laws for opto-electronics and cavity options could be used further to reduce energy consumption of the modulators performing the binary-weighting of PBW-DAC [41, 42].

**Acknowledgements**
V.S. is supported by the Air Force Office of Scientific Research under award number FA9550-19-1-0277, and  by the Office of Navy Research under award number N00014-19-1-2595 of the Electronic Warfare program. The authors would also like to acknowledge useful discussions with Nicholas G. Usechak.

**Author Contributions**
J.G. and V.S. conceived the idea. J.W. fabricated and tested the PIC samples. J.W and M.M. performed the data analysis and simulation results. M.M. and A.B. conceptualizing of the context and application placement. All authors discussed the results and commented on the manuscript.

**Additional Information**
Supplementary information accompanies this paper at …
Competing Interests: The authors declare no competing interests.